\documentclass[letter,longauth,traditabstract]{aa}

\usepackage{amsmath}
\usepackage[varg]{txfonts}
\usepackage{graphicx}

\usepackage{xspace}
\usepackage[utf8]{inputenc} 

\usepackage{url}
\usepackage{xspace}

\newcommand{\XMMNewton}{\textsl{XMM-Newton}\xspace}

\newcommand{\Swift}{\textsl{Swift}\xspace}

\newcommand{\Fermi}{\textsl{Fermi}\xspace}

\begin{document}

\title{TANAMI blazars in the IceCube PeV neutrino fields}
\author{
  F. Krau\ss{}\inst{\ref{affil:remeis},\ref{affil:wuerzburg}}
  \and M.~Kadler\inst{\ref{affil:wuerzburg}}
  \and K.~Mannheim\inst{\ref{affil:wuerzburg}}
  \and R.~Schulz\inst{\ref{affil:remeis},\ref{affil:wuerzburg}}
  \and J.~Tr\"ustedt\inst{\ref{affil:remeis},\ref{affil:wuerzburg}}
  \and J.~Wilms\inst{\ref{affil:remeis}}
  \and R.~Ojha\inst{\ref{affil:nasa_gfsc}, \ref{affil:umd}, \ref{affil:cua}}
  \and E.~Ros\inst{ \ref{affil:mpe},\ref{affil:val},\ref{affil:valobs}}
  \and G.~Anton\inst{\ref{affil:ecap}}
  \and W.~Baumgartner\inst{\ref{affil:nasa_gfsc}}
  \and T.~Beuchert\inst{\ref{affil:remeis},\ref{affil:wuerzburg}}
  \and J.~Blanchard\inst{\ref{affil:alma}}
  \and C.~B\"urkel\inst{\ref{affil:remeis},\ref{affil:wuerzburg}}
  \and B.~Carpenter\inst{\ref{affil:cua}}
  \and T.~Eberl\inst{\ref{affil:ecap}}
  \and P.G.~Edwards\inst{\ref{affil:csiro}}
  \and D.~Eisenacher\inst{\ref{affil:wuerzburg}}
  \and D.~Els\"asser\inst{\ref{affil:wuerzburg}}
  \and K.~Fehn\inst{\ref{affil:ecap}}
  \and U.~Fritsch\inst{\ref{affil:ecap}}
  \and N.~Gehrels\inst{\ref{affil:nasa_gfsc}}
  \and C.~Gr\"afe\inst{\ref{affil:remeis},\ref{affil:wuerzburg}}
  \and C.~Gro\ss{}berger\inst{\ref{affil:mpa}}
  \and H.~Hase\inst{\ref{affil:kart}}
  \and S.~Horiuchi\inst{\ref{affil:csiro2}}
  \and C.~James\inst{\ref{affil:ecap}}
  \and A.~Kappes\inst{\ref{affil:wuerzburg}}
  \and U.~Katz\inst{\ref{affil:ecap}}
  \and A.~Kreikenbohm\inst{\ref{affil:remeis},\ref{affil:wuerzburg}}
  \and I.~Kreykenbohm\inst{\ref{affil:remeis}}
  \and M.~Langejahn\inst{\ref{affil:remeis},\ref{affil:wuerzburg}}
  \and K.~Leiter\inst{\ref{affil:remeis},\ref{affil:wuerzburg}}
  \and E.~Litzinger\inst{\ref{affil:remeis},\ref{affil:wuerzburg}}
  \and J.E.J.~Lovell\inst{\ref{affil:tasman}}
  \and C.~M\"uller\inst{\ref{affil:remeis},\ref{affil:wuerzburg}}
  \and C.~Phillips\inst{\ref{affil:csiro}}
  \and C.~Pl\"otz\inst{\ref{affil:kart}}
  \and J. Quick\inst{\ref{affil:hart}}
  \and T.~Steinbring\inst{\ref{affil:remeis},\ref{affil:wuerzburg}}
  \and J.~Stevens\inst{\ref{affil:csiro}}
  \and D.~J.~Thompson\inst{\ref{affil:nasa_gfsc}}
  \and A.K.~Tzioumis\inst{\ref{affil:csiro}}
}

\institute{
  Dr.~Remeis Sternwarte \& ECAP, Universit\"at Erlangen-N\"urnberg,
  Sternwartstrasse 7, 96049 Bamberg, Germany
  \label{affil:remeis} 
  \and
  Institut f\"ur Theoretische Physik und Astrophysik, Universit\"at
  W\"urzburg, Emil-Fischer-Str. 31, 97074 W\"urzburg, Germany
  \label{affil:wuerzburg}
  \and
  NASA, Goddard Space Flight Center, Greenbelt, MD 20771,
  USA  \label{affil:nasa_gfsc}
  \and
  University of Maryland, Baltimore County, Baltimore, MD 21250,
  USA 
  \label{affil:umd} 
  \and
  Catholic University of America, Washington, DC 20064,
  USA \label{affil:cua}
  \and
  Max-Planck-Institut f\"ur Radioastronomie, Auf dem Hügel 69, 53121
  Bonn, Germany 
  \label{affil:mpe}
  \and
  Departament d'Astronomia i Astrof\'isica, Universitat de Val\`encia,
  C/ Dr. Moliner 50,
  46100 Burjassot, Val\`encia, Spain
  \label{affil:val}
  \and
  Observatori Astron\`omic, C/ Catedr\'atico Jos\'e Beltr\'an no. 2,
  46980 Paterna, Val\`encia, Spain
  \label{affil:valobs}
  \and
  ECAP, Universit\"at Erlangen-N\"urnberg, Erwin-Rommel-Str. 1, 91058
  Erlangen, Germany 
   \label{affil:ecap} 
  \and Departamento de Astronom\'ia, Universidad de Concepci\'on,
  Casilla 160, Chile
  \label{affil:alma}
 \and
   CSIRO Astronomy and Space Science, ATNF, PO Box 76 Epping,
   NSW 1710, Australia
   \label{affil:csiro}
  \and Max-Planck-Institut f\"ur extraterrestrische Physik,
  Giessenbachstraße 1, 85741 Garching, Germany 
  Bonn, Germany 
  \label{affil:mpa}
   \and
   Bundesamt f\"ur Kartographie und Geod\"asie, 93444 Bad K\"otzting,
   Germany
   \label{affil:kart}
   \and
   CSIRO Astronomy and Space Science, Canberra Deep Space
   Communications Complex, P.O.\ Box 1035, Tuggeranong, ACT
   2901, Australia
   \label{affil:csiro2}
   \and
   School of Mathematics \& Physics, University of Tasmania, Private
   Bag 37, Hobart, Tasmania 7001, Australia
   \label{affil:tasman}
   \and
   Hartebeesthoek Radio Astronomy Observatory, Krugersdorp, South
   Africa
   \label{affil:hart}
}

\authorrunning{F. Krau\ss~et al.}
\titlerunning{TANAMI Blazars in the IceCube PeV neutrino fields}
\date{Received 15 May 2014 / Accepted 2 June 2014} 

\abstract{ 
The IceCube Collaboration has announced the discovery of a
  neutrino flux in excess of the atmospheric background. 
  Owing to the steeply falling atmospheric background spectrum, events
  at PeV energies most likely have an extraterrestrial origin.
  We present the multiwavelength properties of the six radio-brightest
  blazars that are positionally coincident with these events using
  contemporaneous data of the TANAMI blazar sample, including
  high-resolution images and spectral energy distributions.
  Assuming the X-ray to $\gamma$-ray emission originates in the
  photoproduction of pions by accelerated protons, the integrated
  predicted neutrino luminosity of these sources is high enough to
  explain the two detected PeV events.} 

\maketitle

\section{Introduction}

The detection of neutrinos at PeV energies in excess of the
atmospheric background reported by the IceCube Collaboration
\citep{Icecube2013a,IceCube2013b} has prompted a quest to
identify their extraterrestrial sources. The two events with PeV
energies (event 20, dubbed `Ernie' and event 14, `Bert', hereafter E20
and E14), detected between May 2010 and May 2012\footnote{A third
  neutrino at 2 PeV (event 35, dubbed ‘Big Bird’) has recently been
  reported for the third year of data (Aartsen et al. 2014).}, have
angular uncertainties of $10\fdg7$ and $13\fdg2$, respectively.

A Galactic center origin has been considered \citep{Razzaque2013}, but
a single source has been excluded by \cite{ANTARES2014}. Pevatrons in
the Galactic center region, such as young supernova remnants, produce
neutrinos at well below 1\,PeV \citep{Aharonian1996}. The 
overall distribution of all 28 IceCube events is consistent with an
isotropic source population, and therefore extragalactic sources are
the prime suspects.   
Neutrino emission has been theoretically predicted from the cores of
active galactic nuclei (AGN) \citep{Stecker2013}, AGN jets
\citep{Mannheim1995}, or gamma-ray bursts \citep{Waxman1997}.
Prevailing models for gamma-ray bursts have recently been excluded as
neutrino sources \citep{Abbasi2012}, and revised models predict much
lower neutrino fluxes than the observed excess \citep{Winter2013}. 
Among the models for a diffuse, isotropic neutrino flux at PeV
energies, only the predicted flux of 
$\sim$$10^{-8}\,\mathrm{GeV}\,\mathrm{cm}^{-2}\,\mathrm{s}^{-1}\,\mathrm{sr}^{-1}$
from AGN jets matches the observed excess flux well
\citep{Learned2000}, although it does not explain the absence of
Glashow-resonance events and the possible gap between 400~GeV and 1~PeV.
AGN jets carry a fraction of the total gravitational energy released
during the accretion of matter onto supermassive black holes. If
observed at a small angle to the line of sight, the emission becomes
relativistically boosted, so the source is classified as a blazar.
Their low-energy, non-thermal radiation stems from synchrotron
emission.
The emission at high energies is explained by hadronic or leptonic models.
In hadronic models protons are accelerated and interact with
low-energy photons (e.g., accretion disk) to produce pions \citep[pion
photoproduction,][]{Mannheim1989}. The pion decays and ensuing
cascades generate neutrinos and $\gamma$ rays.
Since the observed spectral energy distributions (SEDs) of AGN result from
the superposition of many emission zones within the jets, the
distinction between hadronic and leptonic emission processes is obscured
by the large number of adjustable parameters.
Unambiguous evidence of hadronic processes could be provided by
neutrino observations. 

In this Letter we address the question of whether the PeV neutrinos
detected by IceCube could originate in blazars by calculating the
expected neutrino fluence. In Sect.~\ref{sec-data}, we describe
multiwavelength data on the six candidate sources. 
In Sect.~\ref{sec:tanami} we present Very Long Baseline Interferometry
(VLBI) images and SEDs and discuss their expected neutrino emission.

\section{Observational data}
\label{sec-data}
Tracking Active Galactic Nuclei with Austral Milliarcsecond
Interferometry
(TANAMI)\footnote{\url{http://pulsar.sternwarte.uni-erlangen.de/tanami/}} 
\citep{2010A&A...519A..45O} is a multiwavelength program that monitors
extragalactic jets of the Southern Sky ($\delta<-30^\circ$).
The sample includes the brightest radio- and $\gamma$-ray (GeV)
blazars. VLBI observations were conducted with the Australian Long Baseline
Array (LBA) in combination with telescopes in South Africa, Chile,
Antarctica, and New Zealand at 8.4\,GHz and 22.3\,GHz.
The DiFX correlator at Curtin University in Perth, Western Australia 
\citep{Deller2007, Deller2011} was used to correlate the data.
Subsequent calibration, hybrid imaging, and image analysis were
performed following \cite{2010A&A...519A..45O}.
TANAMI radio observations are supported by
flux-density measurements with the Australia Telescope Compact Array
(ATCA) \citep{2012arXiv1205.2403S} and the Ceduna 30\,m telescope
\citep{CedunaCul}.

X-ray data taken during the IceCube period are from the TANAMI program and
the public archives of \Swift \citep{Swift2004} and \XMMNewton
\citep{XMM2001} and were supplemented with non-simultaneous archival
data.
\Swift/XRT and \XMMNewton/pn data were reduced with standard methods, using
the most recent software packages (\textsc{HEASOFT
  6.15.1}\footnote{\url{http://heasarc.nasa.gov/lheasoft/}}, 
\textsc{SAS 1.2.}\footnote{\url{http://xmm.esac.esa.int/sas/}}) and
calibration databases. Spectra were grouped to 
a minimum signal-to-noise ratio of 3 for the \Swift/XRT data and 5 for
the \XMMNewton/pn data. Spectral fitting was performed with \textsc{ISIS
  1.6.2} \citep{ISIS2000} using Cash statistics \citep{Cash},
except for the \XMMNewton data, where the higher count rates allow
using $\chi^2$-statistics.  
We fitted the 0.5--10\,keV energy band with an absorbed power-law
model, which yielded good results in all cases. None of the sources
showed evidence of intrinsic X-ray absorption in excess of the
Galactic value \citep{Kalberla}. X-ray data were deabsorbed using
abundances from \cite{2000ApJ...542..914W} and cross sections from
\cite{1996ApJ...465..487V}. \Swift/UVOT and \XMMNewton/OM data were
extracted following standard methods. Optical, infrared, and
ultraviolet data were dereddened using the same absorbing columns
\citep[][and references therein]{2012ApJ...759...95N}. We included
spectral data for these six sources from the \Fermi/LAT
\citep{LAT2009} second source catalog \citep[2FGL]{2fgl}, which
covered the time period 2008 August to 2010 August. We also calculated
spectra for the 2010 May to 2012 May IceCube integration period using
the reprocessed Pass 7 data (v9r32p5) and the P7REP\_SOURCE\_V15
instrumental response functions \citep[IRF;][]{Pass7} and a region of
interest (ROI) of $7^{\circ}$. Non-simultaneous data from the
\Swift/BAT 70-month catalog \citep[][hard X-rays]{Baumgartner2013},
\textsl{Planck} \citep[microwave]{planckearly}, Wide-Field Infrared
Survey Explorer \citep[\textsl{WISE};][]{WISE2010}, and Two Micron All
Sky Survey \citep[2MASS;][]{2MASS2006} (infrared) are also included.
An \textsl{INTEGRAL} \citep{INTEGRAL2003} spectrum has been obtained
for 1653$-$329 for all available data since 2003 using the HEAVENS
online tool \citep{2010heavens}.

\section{Results}
\label{sec:tanami}
\subsection{TANAMI sources in the two PeV-neutrino fields}
Six TANAMI sources are located in the $1\sigma$ positional uncertainty
region for the two PeV events (Table \ref{tab-TanamiSources}). The
three blazars PKS\,B0235$-$618 (in the following referred to as
0235$-$618), PKS\,B0302$-$623 (0302$-$623), and PKS\,B0308$-$611
(0308$-$611) are located in the E20 field. In the E14 field we find
the three blazars Swift~J1656.3$-$3302 (1653$-$329), PMN\,J1717$-$3342
(1714$-$336) and PMN\,J1802$-$3940 (1759$-$396). Of the twelve
brightest $\gamma$-ray sources (in the two fields) from the 2FGL
catalog, only these six named sources have correlated VLBI flux
densities at 8.4\,GHz above 400mJy. All other sources are considerably
fainter with typically 30\,mJy to 160\,mJy at 1.4\,GHz and on kpc
scales \citep{Condon1998}. The source 0235$-$618 is formally also
consistent with IceCube event 7 (34.3\,TeV), while 1653$-$329 and 1714$-$336
are also within the error circles of events 2 (117\,TeV) and 25 (33.5\,TeV).
The source 1759$-$396 agrees with the positions of events 2 (117\,TeV),
15 (57.5\,TeV), and 25 (33.5\,TeV).

\subsection{VLBI images}
\begin{figure*}
\hspace*{0.01\textwidth}\begin{minipage}{0.735\textwidth}
\begin{minipage}{\textwidth}
\includegraphics[width=0.321\textwidth]{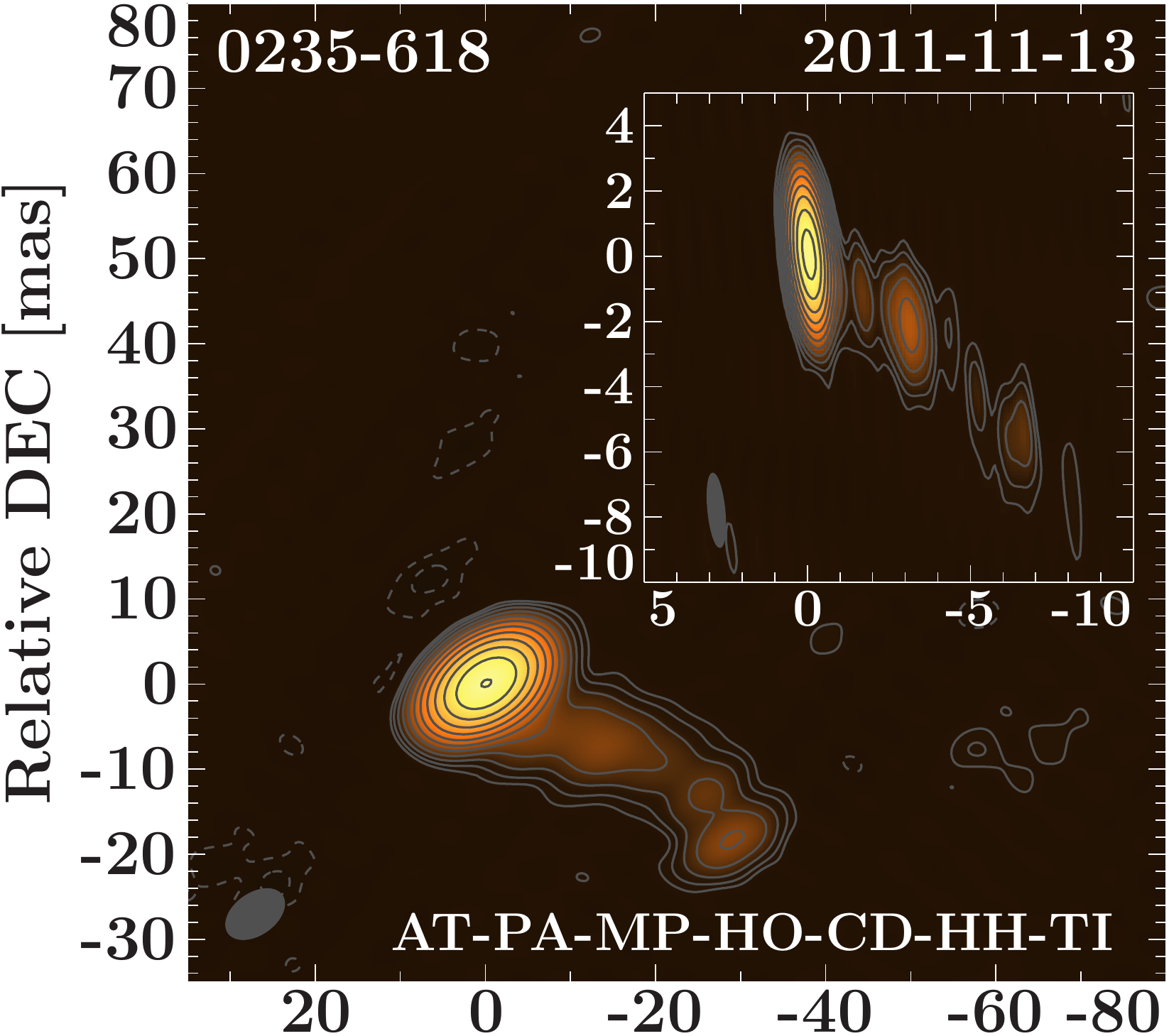}
\includegraphics[width=0.3\textwidth]{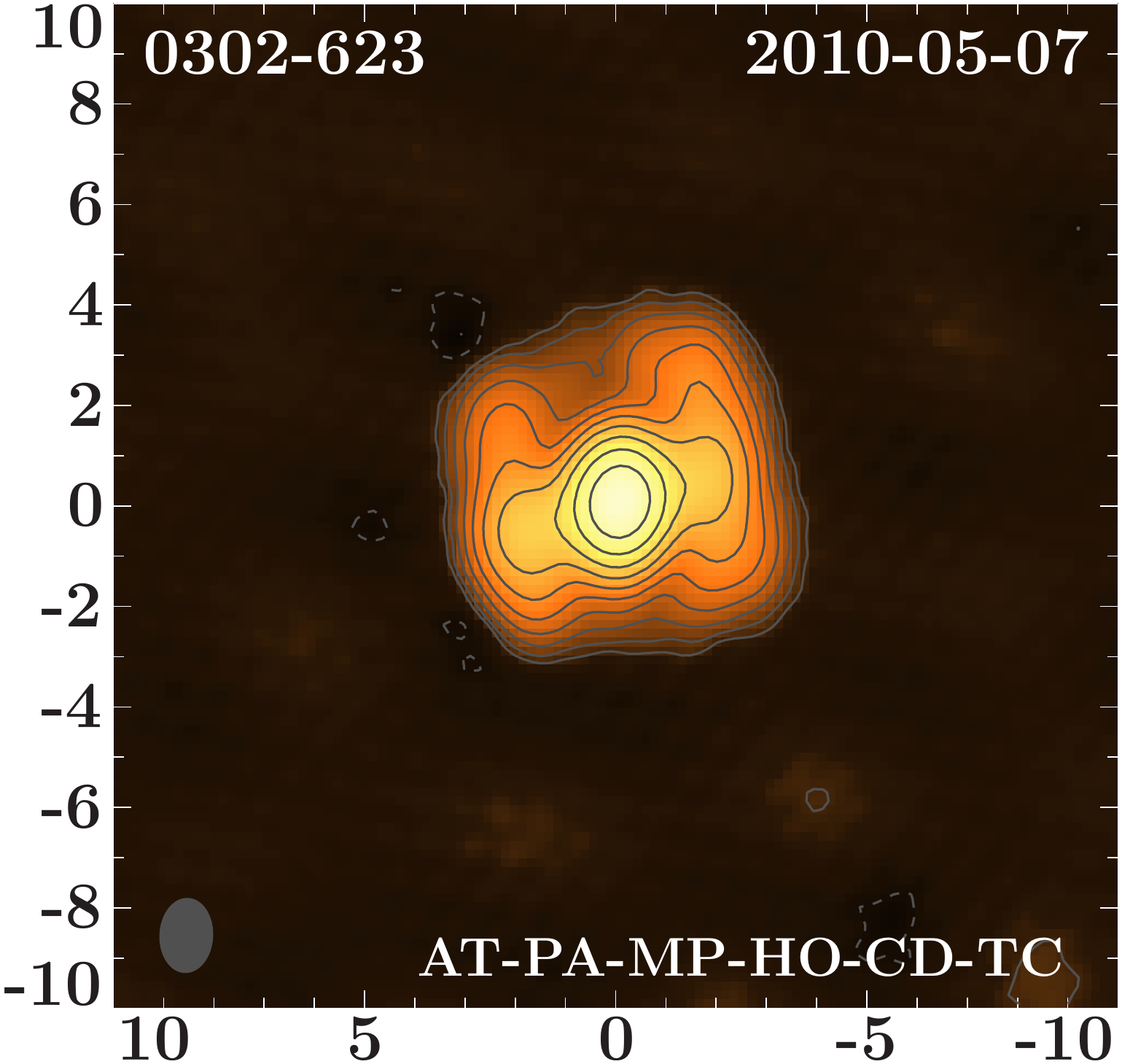}
\includegraphics[width=0.362\textwidth]{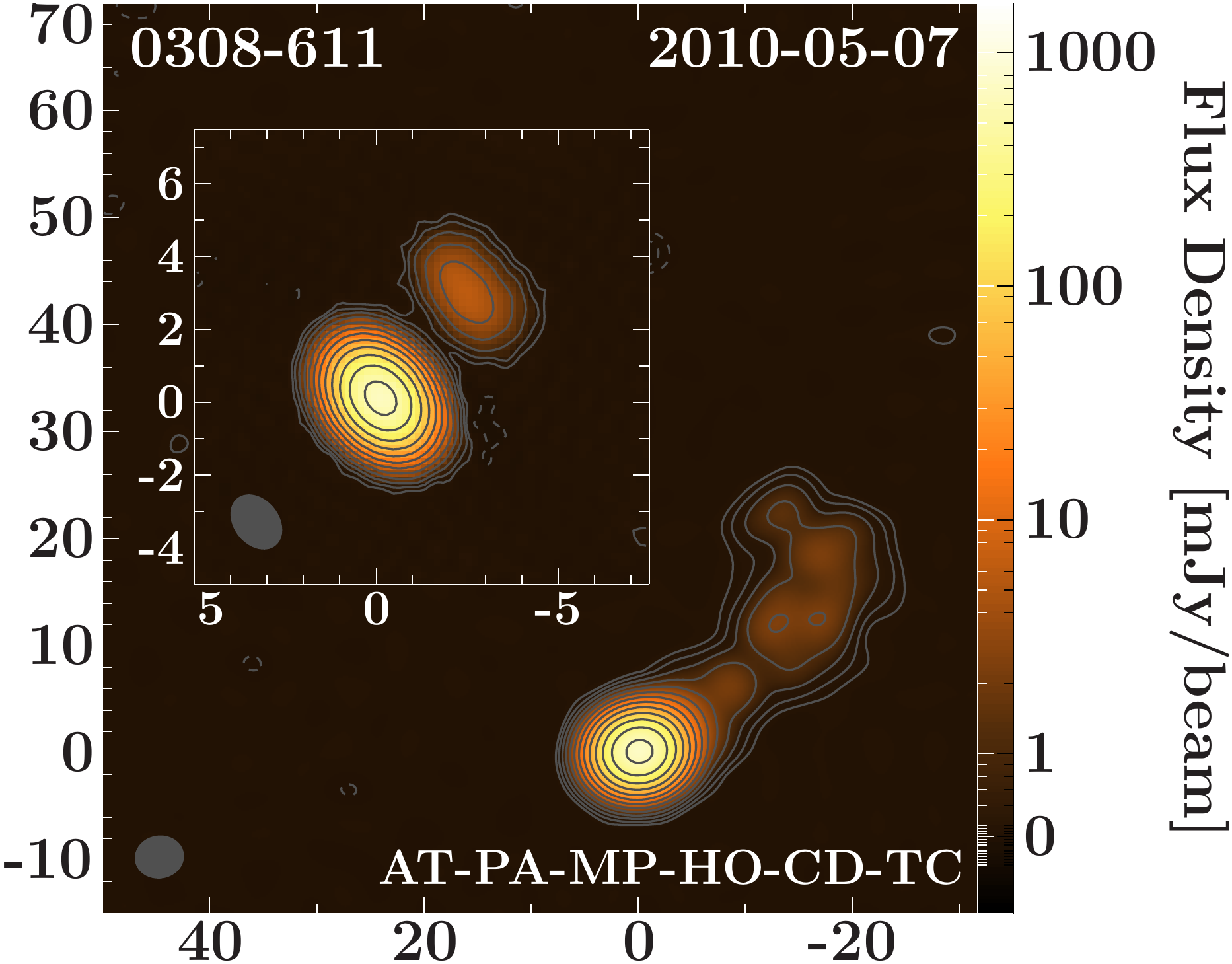}
\end{minipage} 
\begin{minipage}{\textwidth}
\includegraphics[width=.321\textwidth]{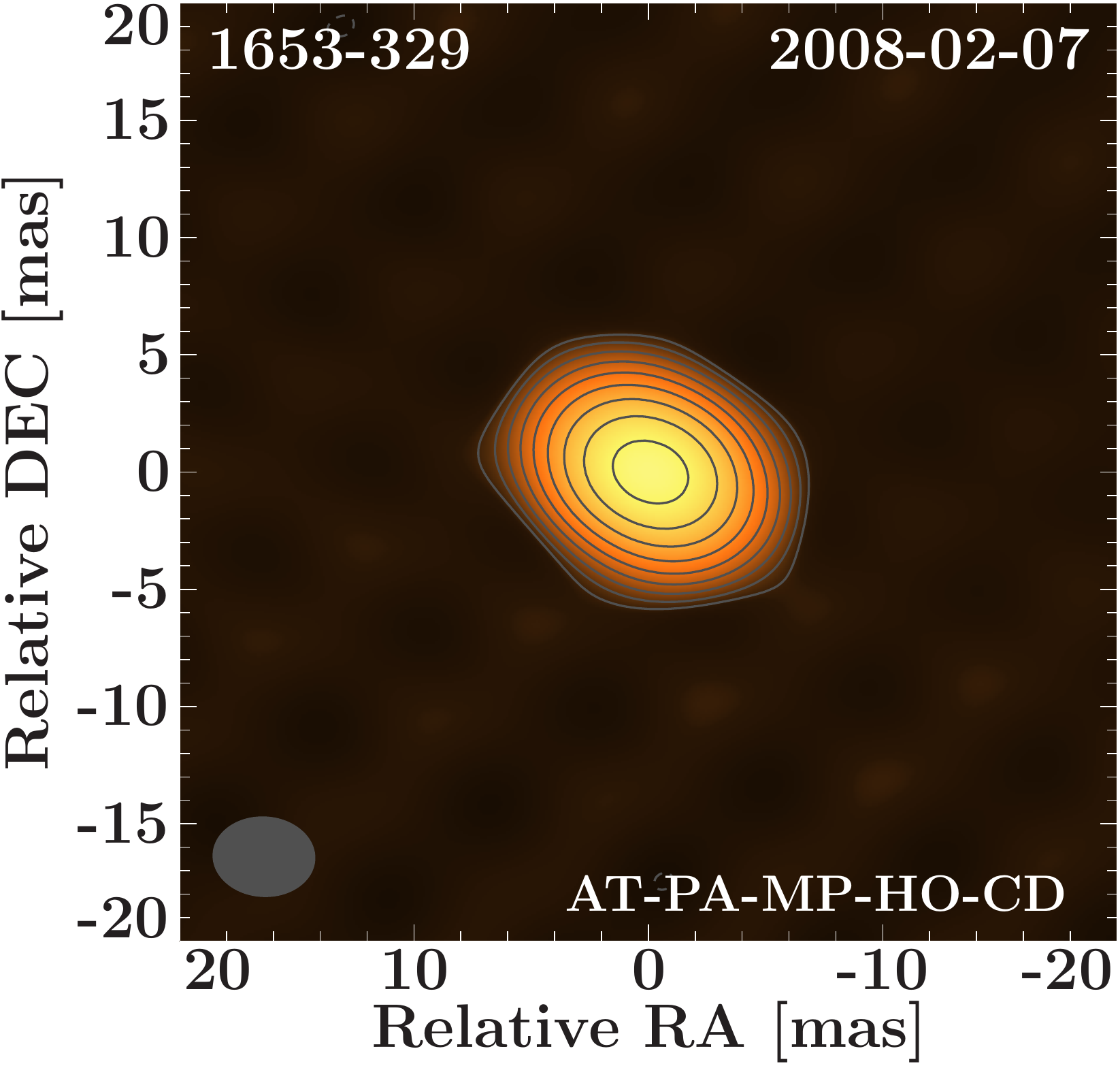}
\includegraphics[width=.3\textwidth]{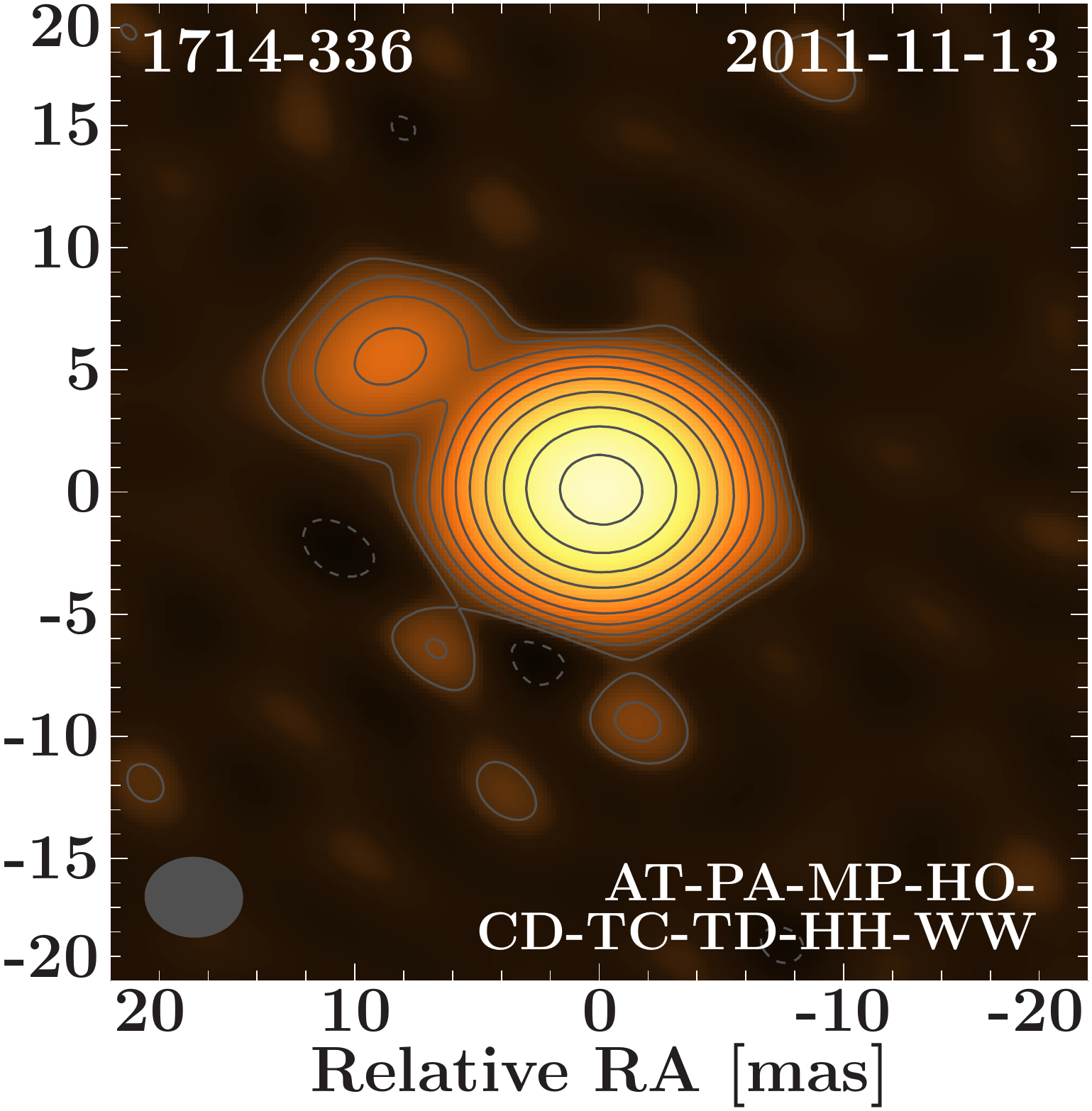}
\includegraphics[width=.362\textwidth]{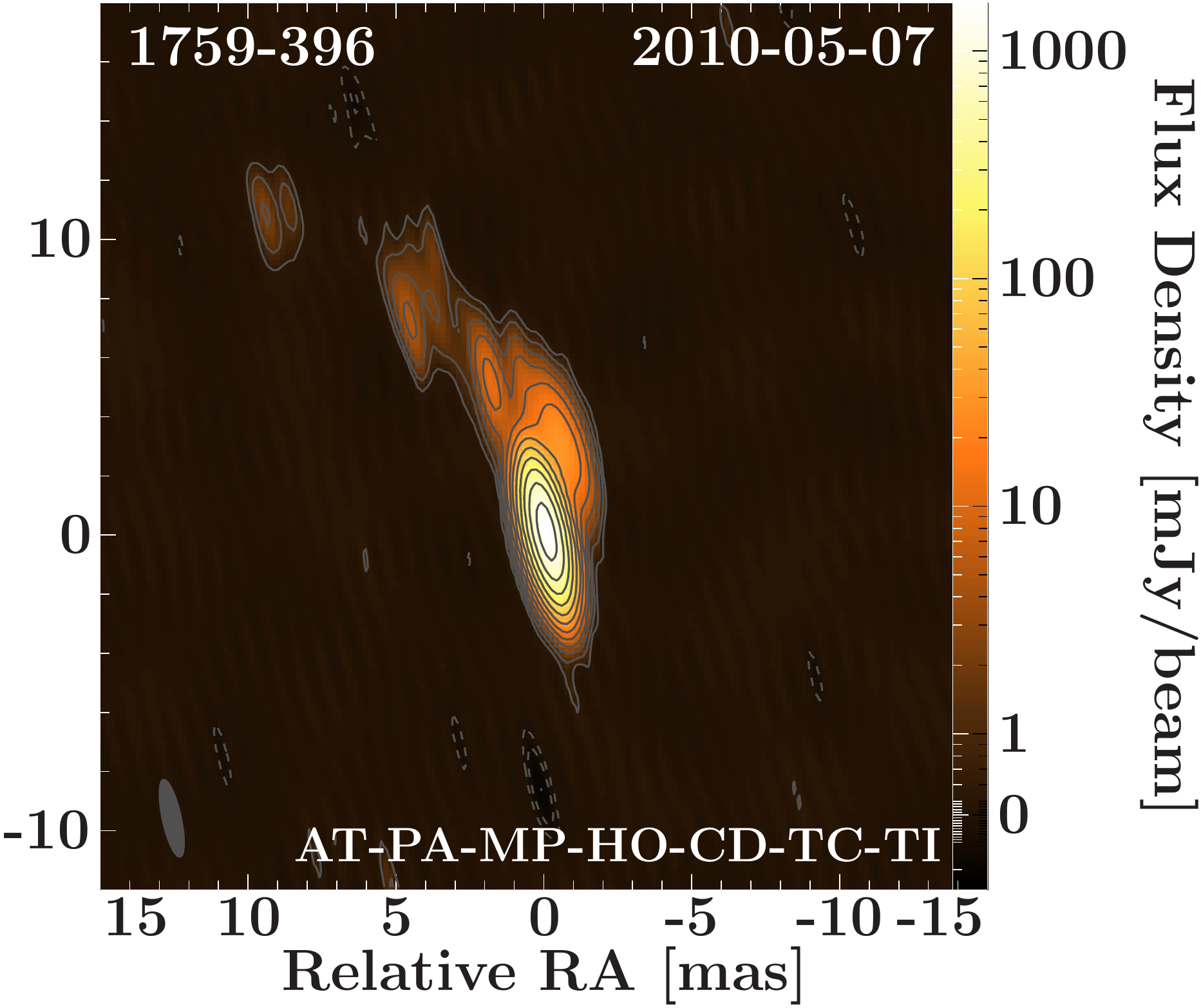}
\end{minipage}
\end{minipage}\hspace*{0.01\textwidth}
\begin{minipage}{0.245\textwidth}
\vspace*{-1.5\baselineskip}
\caption{VLBI images at 8.4\,GHz (natural weighting 
  for insets of 0235$-$618 and 0308$-$611 and all other images; tapered to 
  10\% at 100 M$\lambda$ for main panels of 0235$-$618
  and 0308$-$611). Restoring beams are shown in the bottom left corners. The
  color scale covers the range between 
  the mean noise level and the maximum flux density (1759$-$396, see
  Table~\ref{tab-TanamiObsInfo} for image parameters). Contour lines start
  at $3 \sigma_\textrm{rms}$ and increase logarithmically by factors
  of 2. The array is given in the bottom right corner: PA: Parkes,
  AT: ATCA, MP: Mopra, HO: Hobart,
  CD: Ceduna, HH: Hartebeesthoek, TC: Tigo, TI: Tidbinbilla (70\,m),
  TD: Tidbinbilla (34\,m), WW:Warkworth} 
\label{fig-VLBIimages}
\end{minipage}
\end{figure*}

The TANAMI VLBI jets of 0235$-$618, 0308$-$611, and 1759$-$396 are
one-sided, indicating relativistic boosting at small angles to the
line of sight (see Fig.~\ref{fig-VLBIimages}). The northwest
direction of the 0308$-$611 jet does not agree with the position angle
indicated by the VLBI Space Observatory Program (VSOP) image of
\cite{2008ApJS..175..314D}, which might be due to jet curvature or the
limited $(u,v)$-coverage of VSOP. The source 0302$-$623, which appeared point-like in
\cite{2004AJ....127.3609O}, shows a highly peculiar morphology with a
compact core and a strong halo-like emission region around the core. 
The east-west extension agrees with
\cite{2008ApJS..175..314D}. We find a high brightness
temperature\footnote{We derived brightness temperatures following
  \citet{Kovalev2005} from Gaussian model fits to the visibility
  data.} of several times $10^{11}$\,K in four objects, which is
typical of $\gamma$-ray-emitting blazars \citep{2012Lindford}. 
We find that 1714$-$336 is substantially scatter broadened. 
The image of 1653$-$329 is from one single scan in 2008 February, outside
the IceCube integration period and does not
have the same quality as other TANAMI images.
\begin{figure*}
  \begin{minipage}{\textwidth}
    \includegraphics[height=0.145\textheight]{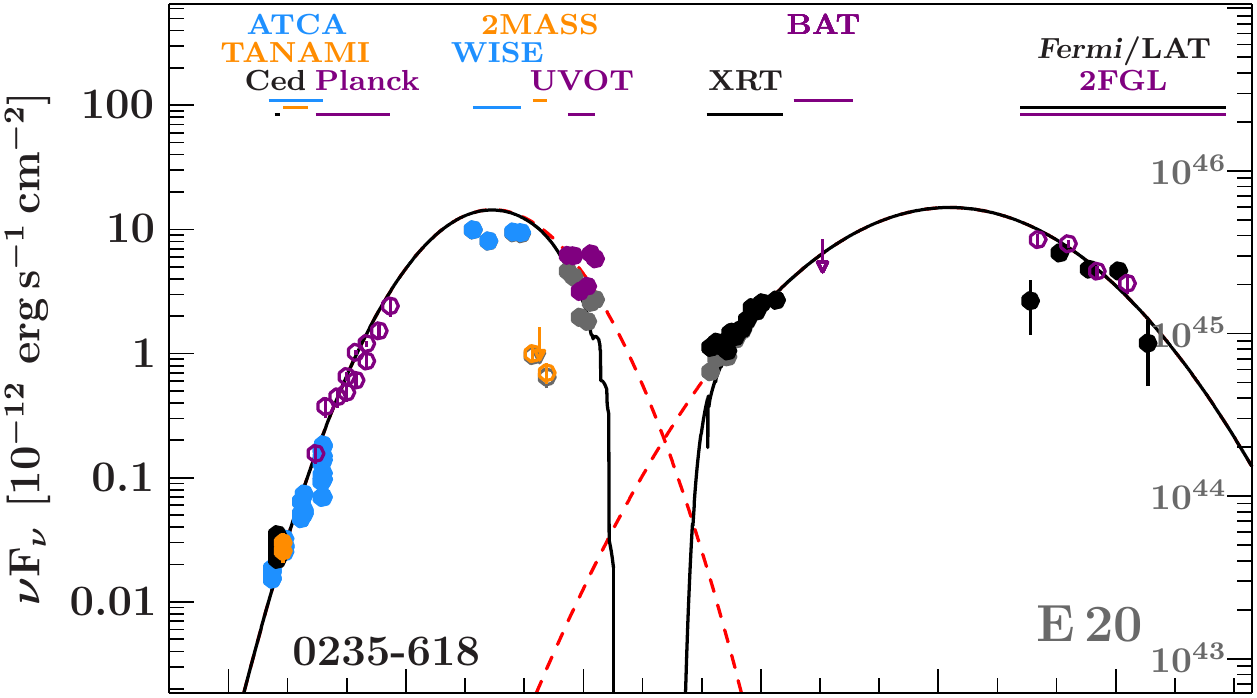}
    \includegraphics[height=0.145\textheight]{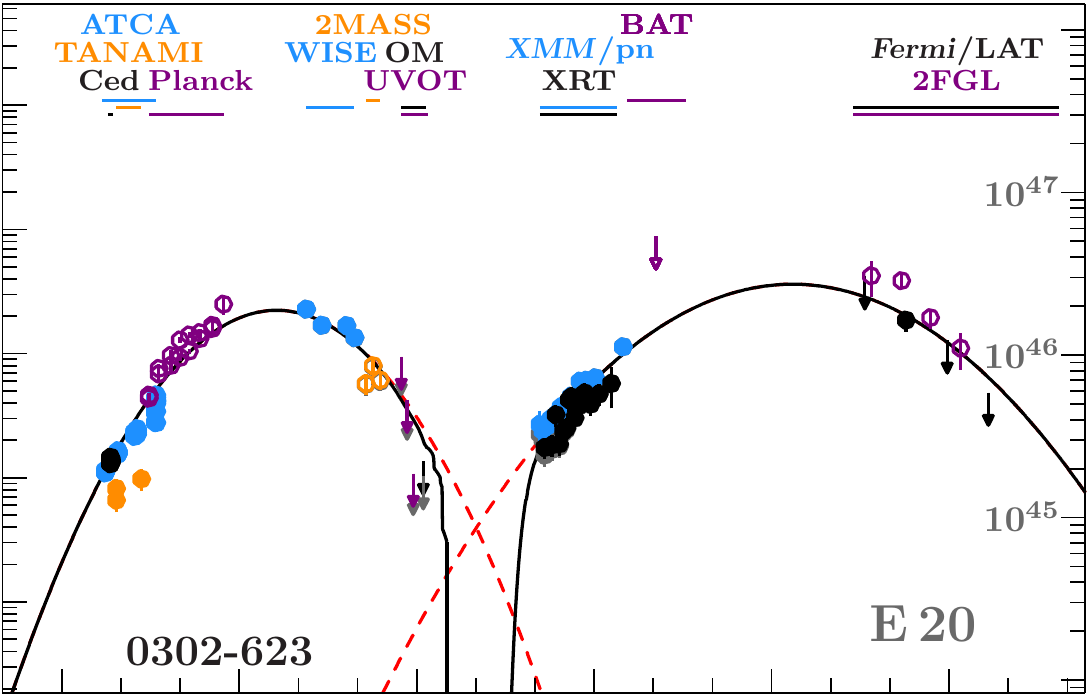}
    \includegraphics[height=0.145\textheight]{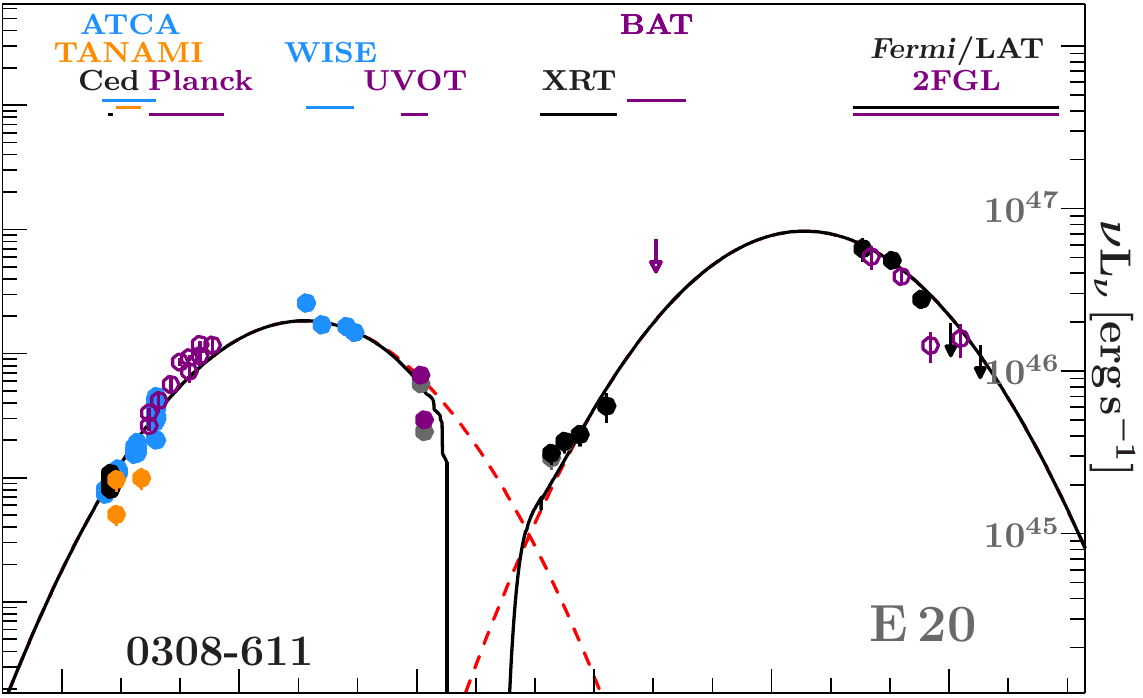}
 \end{minipage}
  \begin{minipage}{\textwidth}
    \includegraphics[height=0.1681\textheight]{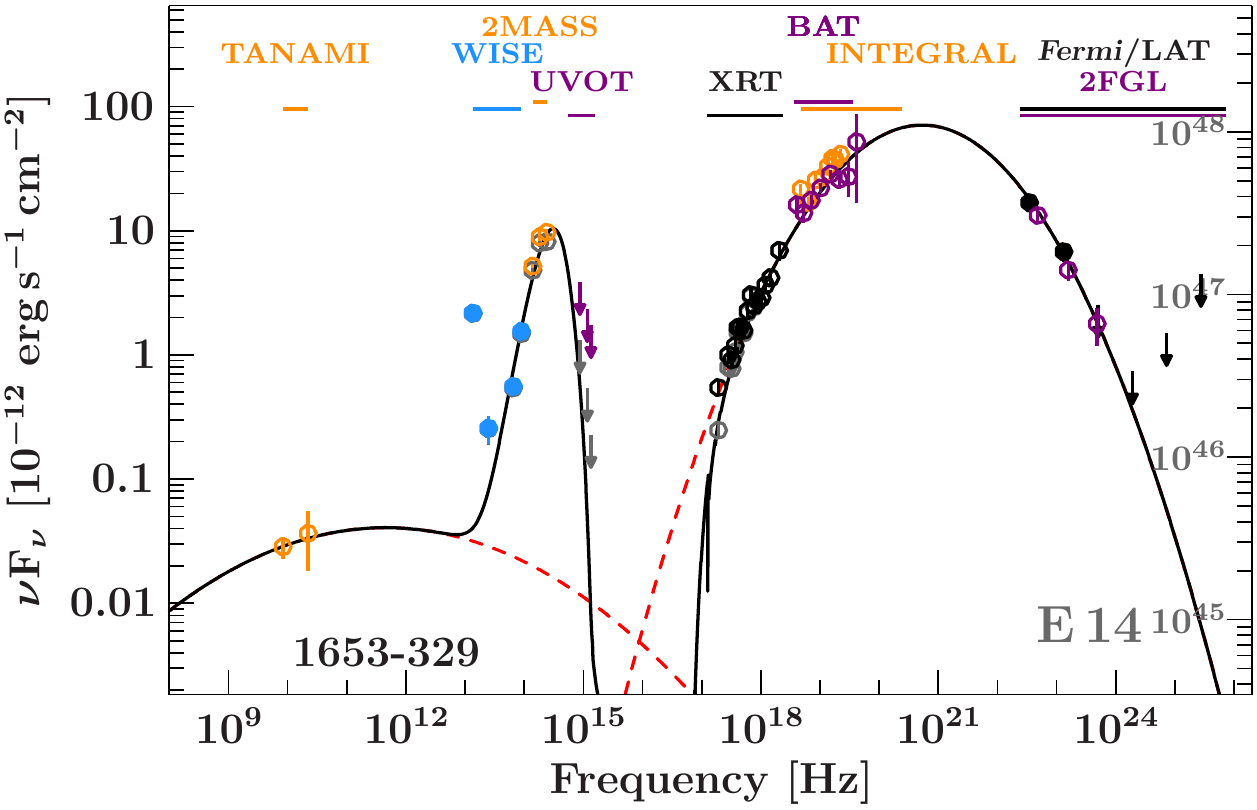}
    \includegraphics[height=0.1681\textheight]{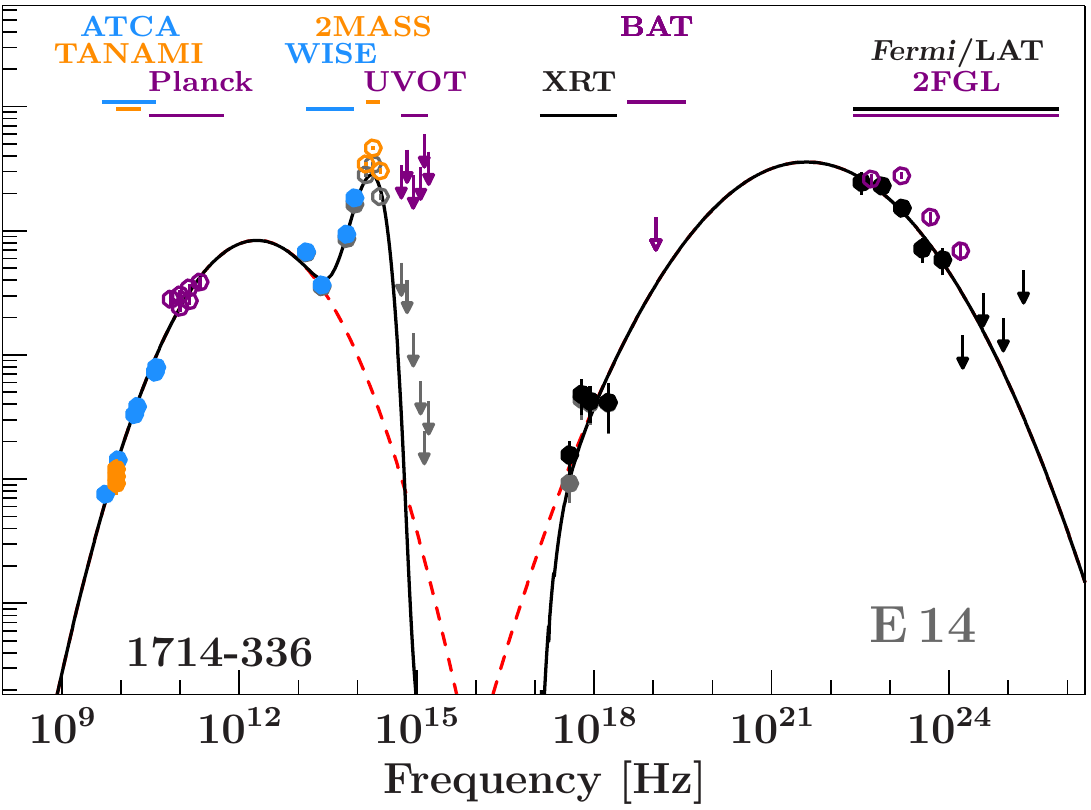}
    \includegraphics[height=0.1681\textheight]{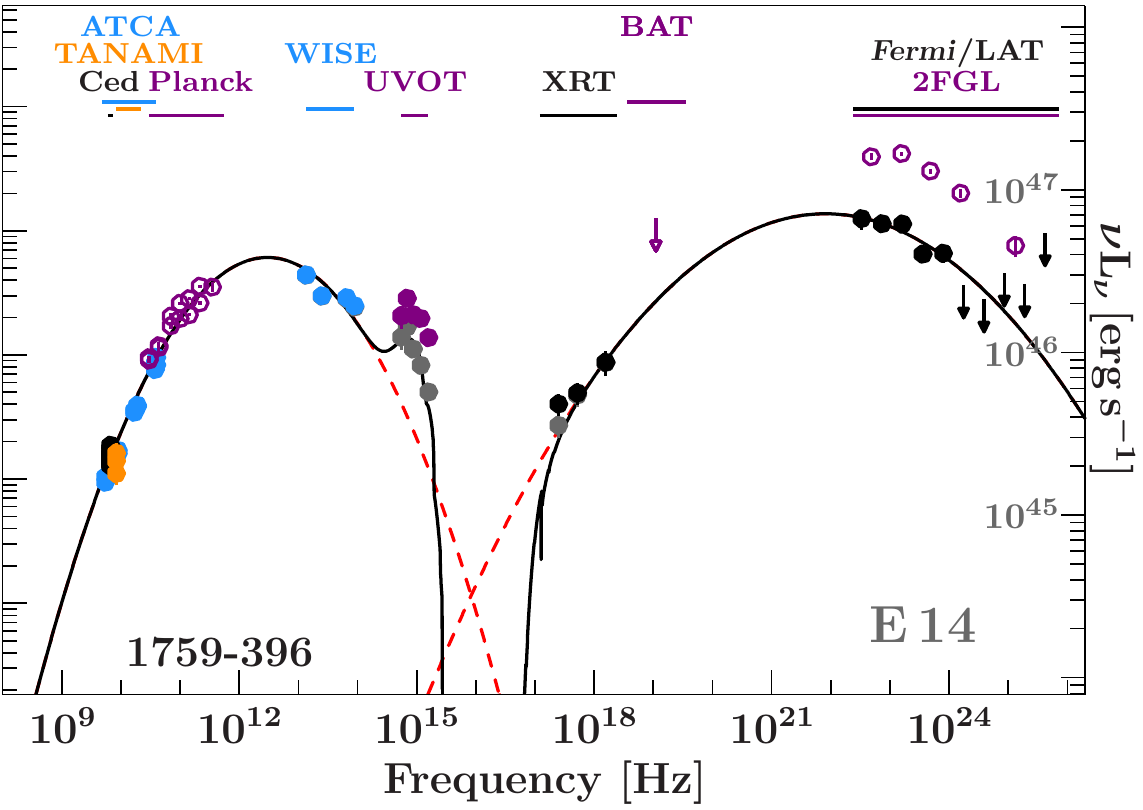}
\end{minipage}
\begin{minipage}{\textwidth}
\caption{Broadband SEDs for the six TANAMI blazars. Filled data points
are from the IceCube integration period (2010\,May--2012\,May), open
circles are archival data outside the time period. Gray shows
the absorbed (X-ray) and reddened (optical/UV) spectra.
The data have been parametrized with logarithmic parabolas (red dashed
lines) modified by extinction and absorption, as well as by an additional
blackbody component where necessary (black solid lines). 
}
\label{fig-sed}
\end{minipage} 
\end{figure*}

\subsection{Broadband spectra}\label{sec:sed}
For all six sources, we find characteristic double-humped
blazar SEDs (Fig.~\ref{fig-sed}). The source 1653$-$329 has an
unusually dominant high-energy hump and is bright at hard X-rays
\citep{2008A&A...480..715M,Baumgartner2013}, while only upper limits 
are placed on the 14--195\,keV flux by \textit{Swift}/BAT for the
other sources, based on the 3$\sigma$ level of background variations
in the survey maps.
The high-energy peak frequencies lie between $10^{20}$\,Hz and
$10^{22}$\,Hz. 1653$-$329 and 1714$-$336, and possibly 1759$-$396
show an additional component between $10^{14}$\,Hz and $10^{15}$\,Hz,
which could be explained by a thermal accretion disk.

\section{Discussion}
\label{sec:discussion}
\subsection{Possible other AGN sources of the IceCube events}
\label{sec:oth}
The six TANAMI blazars are the brightest radio and $\gamma$-ray
emitting AGN in the two IceCube PeV event fields.
The two moderately bright extragalactic radio
sources PKS\,1657$-$261 and PKS\,1741$-$312 (270\,mJy and 470\,mJy
compact flux density at 8.4\,GHz and 8.6\,GHz, respectively) with
compact jets \citep{2004AJ....127.3609O,2005AJ....129.1163P,
  Condon1998} have not shown substantial $\gamma$-ray emission in the
2FGL period. The same is true of several hard X-ray detected blazars
and radio galaxies \citep{Baumgartner2013}. 
Four blazars are slightly outside the uncertainty region of
E14:  
NRAO\,530, PKS\,1622$-$29, PKS\,1622$-$253, and PKS\,1830$-$211 at
$15\fdg0$, $16\fdg8$, $17\fdg3$, and $13\fdg5$ distance to the
coordinates at $13\fdg2$ error radius.

\subsection{Expected neutrino rate from pion photoproduction}

Proton acceleration occurs in blazar jets moving with bulk Lorentz
factor $\Gamma$.
In pion photoproduction the neutrino flux is related to the
bolometric high-energy electromagnetic flux.
We consider the illustrative case of isospin symmetry
(equal numbers of $\pi^+$, $\pi^-$, and $\pi^\circ$). 
We obtain the neutrino flux for the three neutrinos among the four
light final-state leptons in charged pion decays
$F_\nu={{2}/{3}}\cdot {{3}/{4}}\cdot F_\pi =
{{1}/{2}}\cdot F_\pi$, and the $\gamma$-ray flux after accounting
for the conversion of electrons and positrons into $\gamma$ rays by cascading,
$F_\gamma={{1}/{3}}\cdot F_\pi+{{1}/{4}}\cdot {{2}/{3}}\cdot
F_\pi={{1}/{2}}\cdot F_\pi$ and therefore $F_\nu=F_\gamma$.
Monte-Carlo simulations confirm this simple estimate
\citep{Muecke2000}. 
Neutrino oscillations establish full-flavor mixing 
across extragalactic distance scales, and therefore 
$F_{\mathrm{b},\nu_{\rm e}}=F_{\mathrm{b},\nu}/3 = F_{\mathrm{b},\gamma}/3$. 
Electromagnetic cascades emerge at X-ray and $\gamma$-ray energies,
and we approximate the non-thermal bolometric photon flux $F_\gamma$
by the integrated flux between 1~keV and 5~GeV.
The broadband spectra were fit with two logarithmic parabolas
\citep{Massaro2004}, as well as a blackbody component, X-ray
absorption and optical extinction. This fit was then integrated in the
given energy range. 
All but one of the blazars in our sample belong to the FSRQ class,
showing strong emission lines due to photo-ionizing UV light from an
accretion disk that could provide target photons. 
The exception is 1714$-$336, which has been classified as a BL Lac
object \citep{2006A&A...455..773V} but shows a particularly strong big
blue bump in the UV and is possibly a misclassified quasar.  
In the jet's comoving frame (marked with primed quantities), the UV
photons from the disk are redshifted ($\epsilon'=\epsilon/\Gamma$) if
they originate at the base of the jet, or blueshifted
if they come from the outer parts of the disk or are scattered photons. 
Photoproduction of pions starts above the threshold energy $E_{\rm
  p,th}'= 2 (\epsilon'/30\,{\rm eV})^{-1}$~PeV.
The neutrinos carry away $\sim 5\%$ of the proton energy, implying a
neutrino energy of $E_\nu\sim  0.1\Gamma
(\epsilon'/30\,{\rm\,eV})^{-1}$\,PeV in the observer's frame.  
For generic values $\epsilon=30$~eV and $\Gamma=10$, the neutrino
spectrum covers the energy range from $100$\,TeV to $10$\,PeV.
Details of the spectrum, however, are subject to model assumptions and
beyond the scope of this paper. 
IceCube would have measured the following number of electron neutrino events 
$ N_{\nu_{\rm e}}(E_\nu)\simeq A_{\rm eff}(E_\nu) (F_{\nu_{\rm
    e}}/E_\nu) \Delta t$.
Adopting $E_\nu=1\,$PeV as the neutrino-production peak energy, an
exposure time of $\Delta t=662$\,days, and an effective area of $A_{\rm
  eff}=10^5$\,cm$^2$ for contained PeV events, we obtain the values
listed in Table~\ref{tab-events}. 
The numbers would be lower for a realistic spectrum of the emitted
neutrinos or if some fraction of the emission were of a leptonic,
proton-synchrotron, or Bethe-Heitler origin.
The steepness of the blazar $\gamma$-ray luminosity function
\citep{Singal2012}, implies that in a large field, the neutrino
fluence will have significant contributions from the brightest sources
in the field, as well as from fainter, unresolved sources.

\section{Conclusions}
The six candidate sources from the TANAMI sample are the
radio-brightest blazars in the neutrino error fields. Assuming that 
the high-energy emission stems from pion photoproduction due to
accelerated protons, the maximum expected number of electron
neutrino events from the six blazars in 662 days is $1.9\pm
0.4$. This is surprisingly close to the actual number of observed
events, given the additional neutrinos expected from a large number of
remote, faint blazars not included in the TANAMI sample. 
The most promising candidate sources are the three TANAMI blazars in
the E14 field, with the highest predicted neutrino rates and the
prevalence of blue bumps. The detection statistics of
neutrinos at these low fluxes is expected to be Poisson-distributed.
For N=1, the 1$\sigma$ single-sided lower and upper limits are 0.173
and 3.300, respectively. With a predicted neutrino fluence of
0.39/1.55 events for the E20/E14 field, we are well inside the Poisson
uncertainty ranges. The six TANAMI sources alone are already capable
of producing the observed PeV neutrino flux.

\begin{acknowledgements}
  We thank the referee for the helpful comments.
  We acknowledge support and partial funding by the Deutsche
  Forschungsgemeinschaft grant WI 1860-10/1 (TANAMI) and GRK 1147,
  Deutsches Zentrum f\"ur Luft- und Raumfahrt grant
  50\,OR\,1311/50\,OR\,1103, and the
  Helmholtz Alliance for Astroparticle Physics (HAP).
  E.\,R. was partially supported by the Spanish MINECO projects
  AYA2009-13036-C02-02, and AYA2012-38491-C02-01 and by the Generalitat
  Valenciana project PROMETEO/2009/104, as well as by the COST MP0905
  action ``Black Holes in a Violent Universe''.
  We thank J.E.~Davis and T.~Johnson for the development of the
  \texttt{slxfig} module and the SED scripts that have been used to 
  prepare the figures in this work.
  This research has made use of a collection of ISIS scripts provided
  by the Dr. Karl Remeis-Observatory, Bamberg, Germany at
  \url{http://www.sternwarte.uni-erlangen.de/isis/}. 
  The Long Baseline Array and Australia Telescope Compact Array are part
  of the Australia Telescope National Facility, which is funded by the
  Commonwealth of Australia for operation as a National Facility managed
  by CSIRO. 
  The \textit{Fermi}-LAT Collaboration acknowledges support for LAT
  development, operation and data analysis from NASA and DOE (United
  States), CEA/Irfu and IN2P3/CNRS (France), ASI and INFN (Italy),
  MEXT, KEK, and JAXA (Japan), and the K.A.~Wallenberg Foundation, the
  Swedish Research Council, and the National Space Board (Sweden).
  Science analysis support in the operations phase from INAF (Italy)
  and CNES (France) is also gratefully acknowledged. 
 \end{acknowledgements}

\begin{table}
  \centering
  \caption{TANAMI sources compatible with the two IceCube PeV events.
    \label{tab-TanamiSources}}
  \small{
    \begin{tabular}{llllll}
      \hline
      Source & R.A.[$^\circ$] & De.c[$^\circ$] & $z$ &
      Class. & $\Theta$ [$^\circ$]\\
      \hline
      0235$-$618 & 39.2218$^{\triangle}$  &
      $-61.6043^{\triangle}$  & 0.47$^{\blacklozenge}$  &
      FSRQ$^{\blacklozenge}$ & 5.61 \\
      0302$-$623 & 45.9610$^{\dagger}$ & $-62.1904^{\dagger}$ &
      1.35$^{\blacklozenge}$ & FSRQ$^{\blacklozenge}$ & 5.98\\
      0308$-$611 & 47.4838$^{\dagger}$ & $-60.9775^{\dagger}$ &
      1.48$^{\blacklozenge}$ & FSRQ$^{\blacklozenge}$ & 7.39\\
      \hline
      1653$-$329& 254.0699$^{\triangle}$  & 
      $-33.0369^{\triangle}$ & 2.40$^\diamond$ & FSRQ$^\diamond$ & 11.18\\
      1714$-$336 & 259.4001$^\star$ & $-33.7024^\star$ &?& BL
      Lac$^\blacktriangle$ & 7.87\\
      1759$-$396 & 270.6778$^\bullet$  & $-39.6689^\bullet$ &
      1.32$^\blacksquare$ & FSRQ$^\blacksquare$ & 12.50 \\
      \hline
    \end{tabular}
  }
  \tablefoot{Columns:
    (1) IAU B1950 name, (2) right ascension, (3) declination,
    (4) redshift, (5) optical classification, (6) angular distance to
    IceCube event coordinates\\
    $^\blacklozenge$ \cite{2008ApJS..175...97H},
    $^{\triangle}$ \cite{2003yCat.2246....0C},
    $^{\dagger}$ \cite{2009A&A...493..317L},
    $^\star$ \cite{2011ApJS..194...25I},
    $^\bullet$  \cite{2003AJ....126.2562F},
    $^\blacktriangle$ \cite{2006A&A...455..773V},
    $^\blacksquare$  \cite{2009A&A...495..691M},
    $^\diamond$  \cite{2008A&A...480..715M}%
  }
\end{table}

\begin{table}
  \caption{Details of interferometric observations and image parameters}
  \label{tab-TanamiObsInfo}
  \centering
  \small{
    \newlength{\tmpsep}
    \setlength{\tmpsep}{1.1ex}
    \begin{tabular}{c@{\hspace*{\tmpsep}}c@{\hspace*{\tmpsep}}c@{\hspace*{\tmpsep}}c@{\hspace*{\tmpsep}}c@{\hspace*{\tmpsep}}c@{\hspace*{\tmpsep}}c}
      \hline
      Source & $\nu$ &  $S_\mathrm{peak}$\tablefootmark{a} &
     $\sigma_\textrm{rms}$\tablefootmark{a}  &
     $S_\mathrm{total}$\tablefootmark{a} & 
      $T_B$& Beam\tablefootmark{a}\\ 
      \hline
      0235$-$618 & 8.4 & 0.32 & 0.08 & 0.38 & 1.6 & $0.51\times 2.28$, $5.8$ \\
      & & (0.35)& (0.06) & (0.37)& & ($4.85\times 7.70$, $-54.9$) \\
      0302$-$623 & 8.4 & 0.83 & 0.29 & 1.38 &1.9&  $1.05\times 1.47$, $-2.8$ \\
      & 22.3 & 0.45 & 0.12 & 0.69 &1.2&  $1.59\times 2.28$, $87.6$ \\
      0308$-$611 & 8.4 & 0.68 & 0.09 & 0.77 &2.0&  $1.20\times 1.64$, $38.8$ \\
      & & (0.73)& (0.05) & (0.77) &&  ($3.89\times 4.49$, $-80.9$) \\
      & 22.3 & 0.50 & 0.13 & 0.54 &0.3&   $1.53\times 1.82$, $-75.9$\\
      1653$-$329 & 8.4 & 0.28 & 0.26 & 0.34 &0.1&  $3.38\times 4.33$, $86.8$ \\
      & 22.3 & --\tablefootmark{b} & --\tablefootmark{b} & 0.17\tablefootmark{b} &&  --\tablefootmark{b} \\
      1714$-$336 & 8.4 & 0.74 & 0.36 & 1.27 &0.02\tablefootmark{c}&  $3.26\times 3.98$, $87.8$ \\
      1759$-$396 & 8.4 & 1.63 & 0.18 & 2.01 &3.1&  $0.64\times 2.70$, $12.2$ \\
      & 22.3 & 1.12 & 0.18 & 1.19 &0.2&  $1.47\times 4.32$, $78.4$\\
      \hline
    \end{tabular}
  }
  \tablefoot{\footnotesize
    Columns: (1) IAU B1950 source name, (2) observing frequency in GHz,
    (3) peak flux density in $\mathrm{Jy/beam}$, (4) image noise
    level in $\mathrm{mJy/beam}$, (5) total flux density in Jy
    (uncertainties are $\lesssim 10$\,\% and $\lesssim 20$\,\% at
    8.4\,GHz and 22.3\,GHz), (6) minimum core brightness temperature
    in $10^{11}$\,K and (7) restoring beam (size, position angle) in
    mas$^2$ and degree. 
    \tablefoottext{a} {Values in brackets denote the application of
      a Gaussian taper to the visibility data of 10\,\% at a
      baseline length of $100\mathrm{\,M\lambda}$.} 
    \tablefoottext{b} {One baseline experiment, flux density only
      accurate to $\sim$50\%.} 
    \tablefootmark{c} {$z=0$ assumed, affected by interstellar
      scattering broadening.} 
  }
\end{table} 
\begin{table}
\caption{Integrated electromagnetic energy flux from 1~keV to 5~GeV and
expected electron neutrino events at 1~PeV in 662 days of IceCube data
for the six candidate blazars. Errors are statistical only.}
\label{tab-events}
\centering
{\renewcommand{\arraystretch}{1.5}
\begin{tabular}{lcc}
\hline
Source & $F_\gamma(\mathrm{erg}\,\mathrm{cm}^{-2}\,\mathrm{s}^{-1}$) &  events \\
\hline
0235$-$618 & $\left(1.0^{+0.5}_{-0.5}\right)\times 10^{-10}$ &  $0.19^{+0.04}_{-0.04}$ \\ 
0302$-$623 & $\left(3.4^{+0.7}_{-0.7}\right)\times 10^{-11}$ &  $0.06^{+0.01}_{-0.01}$ \\
0308$-$611 & $\left(7.5^{+2.9}_{-2.9}\right)\times 10^{-11}$ &
$0.14^{+0.05}_{-0.05}$ \\
\hline
1653$-$329 & $\left(4.5^{+0.5}_{-0.5}\right)\times 10^{-10}$ &  $0.86^{+0.10}_{-0.10}$ \\
1714$-$336 & $\left(2.4^{+0.5}_{-0.6}\right)\times 10^{-10}$ &  $0.46^{+0.10}_{-0.12}$ \\
1759$-$396 & $\left(1.2^{+0.3}_{-0.2}\right)\times 10^{-10}$ &  $0.23^{+0.50}_{-0.40}$ \\
\hline
Total & & $1.9\pm0.4$\\
\hline
\end{tabular}}
\end{table}

\end{document}